\documentclass[a4paper,twoside]{article}
\newcommand{\affil}[1]{$^{\rm #1}$}
\usepackage[authoryear]{natbib}
\bibpunct{(}{)}{;}{a}{}{,}
\usepackage{graphicx}
\usepackage{url}
\date{} 
\newcommand{\splot}{\textsc{s2plot }}
\newcommand{\splotns}{\textsc{s2plot}}

\title{\large\bf\flushleft Interchanging Interactive 3-d Graphics for 
Astronomy\footnote{Research undertaken as part of the Commonwealth Cosmology Initiative (CCI: www.thecci.org), an international collaboration supported by the Australian Research Council}} 
\author{\parbox{\textwidth}{\flushleft
\vspace{-0.5cm}
{\it C.~J.\ Fluke\affil{A,B}, D.~G. Barnes\affil{A} and N.~T.\ Jones\affil{A}} \\
\vspace{0.4cm}
{\small \affil{A}\,Centre for Astrophysics and Supercomputing, Swinburne University of Technology, PO Box 218, Hawthorn, Australia, 3122}\\
{\small \affil{B}\,Email: cfluke@swin.edu.au}}}
\begin{document}
\twocolumn[
\begin{changemargin}{.8cm}{.5cm}
\begin{minipage}{.9\textwidth}
\vspace{-1cm}
\maketitle
%
%
\small{\bf Abstract:}
We demonstrate how interactive, three-dimensional (3-d) scientific 
visualizations can be efficiently interchanged between a variety of 
mediums. Through the use of an appropriate interchange 
format, and a unified interaction interface, we minimize the effort
to produce visualizations appropriate for undertaking knowledge 
discovery at the astronomer's desktop, as part of conference 
presentations, in digital publications or as Web content.
We use examples from cosmological visualization to address some of the issues of
interchange, and to describe our approach to adapting \splot 
desktop visualizations to the Web.  

\vspace{0.2cm}

Supporting Web examples are available from 
\url{http://astronomy.swin.edu.au/s2plot/interchange}.




\medskip
\medskip
\end{minipage}
\end{changemargin}
]
\small

\section{Introduction}
There are four main ways that astronomers (and scientists in other
disciplines) use visual representations of datasets:
\begin{enumerate}
\item For knowledge discovery;
\item For formal publication of research results;
\item For academic presentations; and
\item For (formal) education and (informal) public outreach.
\end{enumerate}

One of the challenges of developing effective visualizations 
appropriate for all of these targets is the increasingly multi-dimensional
nature of astronomy data.  Consider the wealth of information
offered by four-dimensional radio data cubes of neutral hydrogen clouds 
(RA, Dec., velocity and intensity);
seven-dimensional cosmological simulation outputs (three-dimensional positions 
and velocities, particle mass); 
or the 77-dimensional Hipparcos Main Catalogue \citep{perryman97}.

A wide range of software packages, including custom codes 
and commercial solutions, are available for the steps of processing, 
analyzing and visualizing multi-dimensional astronomy data.
However, despite the wealth of tools and techniques, most of 
astronomy is still performed, presented, taught and learnt in only two dimensions.
Aside from the obvious explanation that producing and displaying 
two-dimensional (2-d) images from scientific datasets is relatively 
straightforward \citep[see item 14 in][]{globus94},
another factor is relevant: 
2-d plots and figures produced for one medium (eg. paper) are easily 
transferable to another medium (eg. screen).  That is, they are easy 
to {\em interchange}. A cartesian scatter plot can be sketched on a whiteboard, 
drawn accurately on-screen for analysis, be saved in suitable format(s) 
for presentation in a lecture or conference talk, and for publication 
in a printed journal. The audience ``uses'' the 2-d plot in the same 
way, regardless of the medium via which it is delivered.


If we extend the visual dimensionality from two to three, no such simple 
interchange mechanism exists.  The visualization tool 
used by the researcher on the desktop is typically not the same one that is 
used to make a conference presentation,  and the creation of educational 
versions of datasets is usually achieved
by handing the data to an animator to interpret using a commercial
animation package. At each
stage, data is converted for use from one tool to another, each step
taking it further away from the original desktop experience, and
requiring significant additional effort for the astronomer. 
Furthermore, there is no common user interface --- a fully interactive 
desktop application is most often reduced to a series of still images. 
What hope, then, for more complex, multi-dimensional datasets?

We have previously reported on the \splot interactive graphics library 
for creating three-dimensional (3-d) visualizations for knowledge
discovery \citep{barnes06}, and demonstrated the novel
capability to produce interactive 3-d figures in the Adobe Portable
Document Format (PDF) to enhance academic publications \citep{barnes08}
and education \citep{fluke08}.
We now describe solutions for interchanging 3-d figures amongst the remaining 
output targets, as shown in Figure \ref{fig:system}. Specifically:

\begin{figure*}
\centering
\includegraphics[width=5.0in]{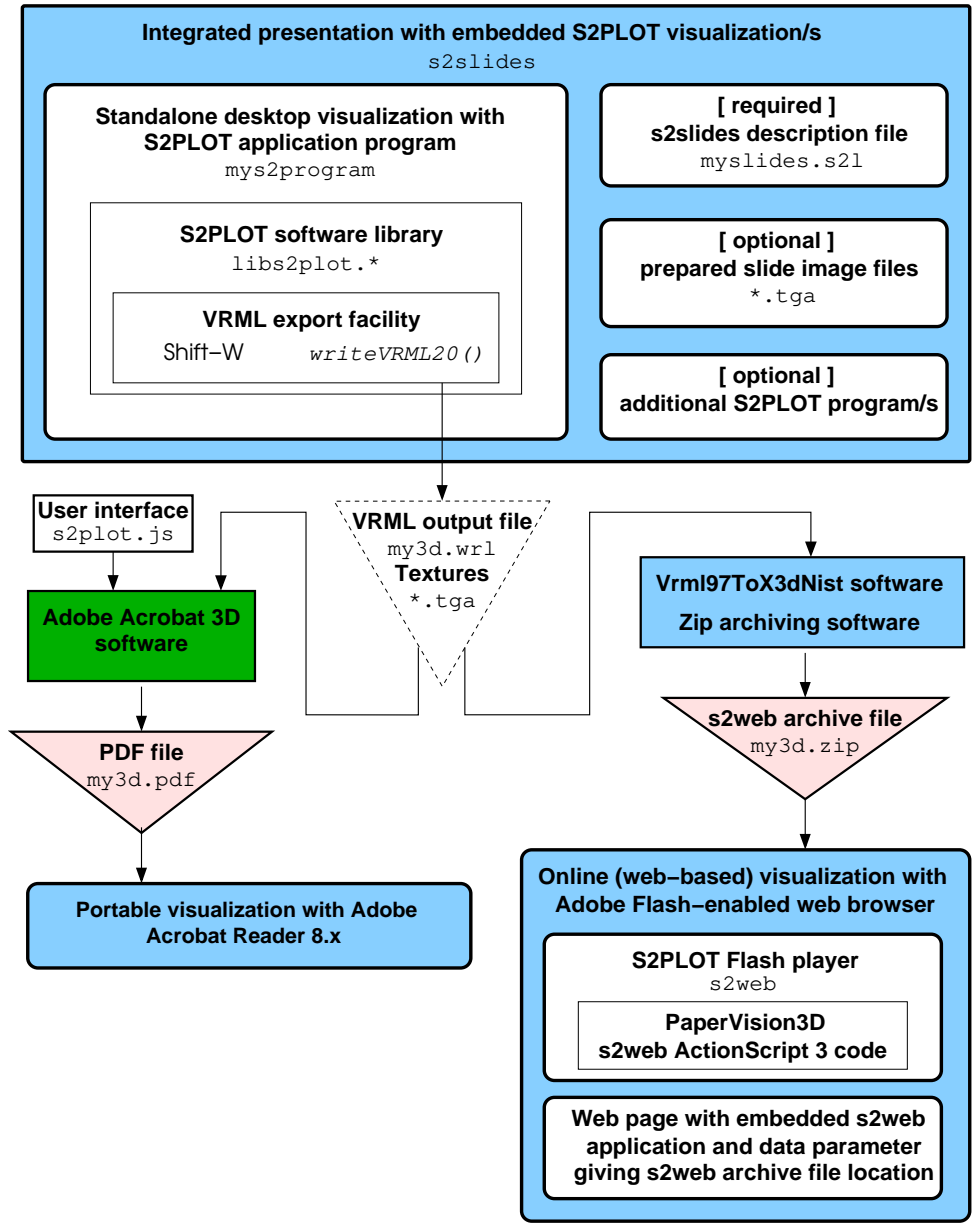} 
\caption{\label{fig:system} An integrated system for creating,
interchanging and using interactive 3-d figures in astronomy, based on
the \splot library.  The large upper box encapsulates desktop
visualization and \textsc{s2slides} presentation modes, while the
lower part of the figure shows the components concerned with
production of 3-d PDF and web-based visualizations.  Shaded triangles
indicate primary output products, and the dashed triangle indicates
the (transient) interchange step: the VRML output file and textures do
not need to be retained once the primary output product/s are created.
Adobe Acrobat 3D is a commercial product.}
\end{figure*}

\begin{itemize}
\item Desktop visualization is accomplished with an application 
(C, C++, FORTRAN, Python; {\tt mys2program} in Figure
\ref{fig:system}) that uses the \splot library. With a programming
interface similiar to the popular \textsc{pgplot} 2-d
library,\footnote{\url{http://www.astro.caltec.edu/~tjp/pgplot}}
\splot provides a straightforward way for astronomers to produce 3-d plots
on-screen through simple function calls within their own custom codes.

\item 3-d figures are embedded in PDF files by exporting the
\splot program geometry to VRML (Virtual Reality Modeling Language;
keypress {\tt Shift-W} or function call {\tt writeVRML20()} in
standard \splot programs), then importing the VRML to Adobe Acrobat 9
Pro Extended, Adobe Acrobat 3D or the Adobe 3D Toolkit (commercial
software).  The standard \splot interaction mode (as well as support
for volume rendering, billboards and so on) is implemented in the
provided {\tt s2plot.js} script, which can be added to the 3-d PDF
model.  The exported VRML file is an interchange file only; it need not 
be kept after the PDF file is created.

\item PowerPoint-style presentation is enabled by the
\textsc{s2slides} application included with \splotns:
it provides for navigation through a sequence of pre-rendered slides
({\tt *.tga} file/s in Figure \ref{fig:system}; e.g. bitmap images
exported from Microsoft PowerPoint), and can overlay stand\-alone
interactive \splot programs on slides according to simple instructions
in an input text file ({\tt myslides.s2l} file in Figure \ref{fig:system}).
Note that
\textsc{s2slides} does {\em not}\/ require Microsoft PowerPoint or any
other commercial presentation software, although such software may be
a convenient way to generate non-interactive slides for presentation
with \textsc{s2slides}.

\item 3-d figures are deployed on the Web by exporting to VRML,
as for the PDF case, converting the VRML to X3D (a simple, syntax-only
conversion accomplished with VrmlToX3dNist). This X3D is then zipped
up with the required textures.  The resultant archive (an ``\textsc{s2web}
archive file'') can then be provided to the
\textsc{s2web} program, which is an Adobe Flash (hereafter ``Flash'') program
that renders \splotns-originated X3D content. \textsc{s2web} allows
the user to control the camera and rendering properties just as they
do for desktop \splot programs.
\end{itemize}

A key feature of our approach to interchange is the preservation of the
user experience, particularly via the keyboard and mouse controls for
interaction, data selection and control of the visual appearance (such 
as rendering mode).

This paper is organized as follows.  
In section 2, we discuss our choice of Web technologies for 
integration with \splot. We detail the practical application of 
these technologies that enable \splot programs to produce 3-d 
visualizations for multimedia deployment.  We further explore the 
motivation for, and implementation of, a common user interface across 
the output targets for \splot programs.  
In Section 3, we consider the specfic case of cosmological data as 
motivation for the demands and requirements of interchanging 3-d 
data and models, and present several applications enabled by our
approach in Section 4.  A brief discussion of our future plans follows in  
Section 5, and we make some concluding remarks in Section 6.

\section{Interchange}
Astronomers have been willing users of Web technologies, most
significantly in the way that they communicate and collaborate
across the world. Astronomers have also taken to fully online, digital
publication of research papers and data archives. Of particular importance
were the first online edition of the Astrophysical Journal
published in September 1995 \citep{abt03}, the abstract searching capabilities
of the ADS Abstract Service \citep{murray92, kurtz00},
the widespread availablity of scanned articles \citep{eichhorn94},
and the establishment of the Centre de Donnees
astronomiques de Strasbourg (CDS) online data archives \citep{genova96},
the precursor to the Virtual Observatory \citep{quinn04}.
Amongst the physical sciences, astronomers are second only
to the particle physics community as users and drivers of Internet
technologies.  

The present problem we are addressing is the provision of a pathway
for publishing and sharing interactive 3-d datasets on the Web.
Accordingly, our approach is to develop and implement the necessary
(informal) standards that yield a practical mechanism for the
creation, interchange and use of interactive 3-d figures. 

\subsection{Exporting {\sc s2plot} Geometry}
\label{sec:vrml}

Simple and robust standardized 3-d interchange formats have been
available for quite some time, notably ISO/IEC 14772:1997 {\em Virtual
Reality Modeling Language} (VRML) and its successor ISO/IEC 19775:2004 
{\em Extensible 3D} (X3D).  When implementing our 3-d PDF solution \citep{barnes08},
it was necessary to choose an
interchange format for exporting 3-d geometry from \splot programs.
We selected VRML because it is one of the few open, non-proprietary
file formats that Adobe Acrobat 3D can import.  Programs using
versions 2.0 and upwards of the \splot library can now write the
\splot geometry data to a VRML file that can be imported into
Acrobat 3D for the creation of 3-d annotations in PDF documents.  

It is worth making the distinction between our use of VRML and
earlier uses of VRML for astronomy visualization 
\citep{crutcher98,plante99,beeson04}.  In previous systems, VRML
was the final product: it had to be viewed in a VRML browser such as
Cortona or FreeWRL.  Historically VRML browsers have not been
particularly complete (in terms of implementation of the entire
language), fully cross-platform or especially stable.  Nor have their
user interfaces been adaptable to suit users' expectations. Our use of
VRML is as an {\em interchange}\/ format: it is simply an intermediate
format for expressing \splot geometry for onward conversion and/or
import into other software.

The scene geometry in VRML is structured as a
model tree --- a concept common to many 3-d applications and scene
description formats.  The tree allows related components of the
geometry to be grouped in a ``branch''.  This tree is preserved upon
VRML import to Acrobat 3D, and Adobe Reader provides both an application
programming interface (API; via Adobe 3D JavaScript) and a simple 
user interface for switching parts of the model
tree on and off.  For our 3-d PDF solution, we
added a function to \splot that allows the user to arrange the
geometry into a tree when exporting VRML.  Selective use of
this function ({\tt pushVRMLname}) makes it possible, for example, to
export two or more different visualizations of the same data from one
\splot program into one VRML file (we demonstrate an example of this
in Section \ref{sct:animation}).  Importing this VRML file to
Acrobat 3D together with some very basic JavaScript code enables
{\em reader}\/ control of the visualization state via buttons or
other user interface elements in the target PDF file.

\subsection{Web Deployment with Flash}

Flash applications are ubiquitous on the Web, which, together
with the breadth and depth of the ActionScript 3 API, makes Flash the ideal system for the
development of a Web-based ``player'' of \splot content.  Unlike
Java---the only other viable environment for Web-based 3-d at this
point in time---Flash applications run in-browser, integrate
seamlessly with static Web content, and the necessary plugin(s) are
generally kept up-to-date.  Ideally for the academic community, Flash
applications can be developed with the freely-available Adobe Flex 3
SDK.

Writing a complete 3-d engine for Flash would be a formidable task.
Happily there already exists a maturing, open-source project
implementing a 3-d environment in Flash:
Papervision3D.\footnote{Papervision3D:
\url{http://blog.papervision3d.org}} Papervision3D is a software 3-d
API for Flash, distributed under the MIT license. It provides a
high-level 3-d engine capable of displaying a standard set of 3-d
primitives, mapping a variety of textures onto objects, including
image and movie textures, and navigating and manipulating 3-d space.
Papervision3D objects can be defined in ActionScript 3 code or can be
read from external files which contain the geometry information.

Papervision3D allows camera and environment redefinition such that we
were able to port the code controlling the \splot interface to
ActionScript 3, giving a very closely matched interface to the one
presented by native \splot desktop applications.  Papervision3D even 
allows for object interaction, so that the \splot {\em interactive}\/
``handles'' can be made available within Flash applications.

Our Web-based solution for 3-d export from \splot entails a simple,
syntactic translation of the exported VRML to X3D.  This translation
retains the model tree structure, and consequently our Flash \textsc{s2web}
program can also provide explicit or implicit user-control of the
model tree and visualization as has been accomplished for PDF.

Predefined strings within the model tree are used to denote \splot 
``handles'' (which are selectable elements of scene geometry) 
and ``billboards'' (textured rectangles which always face the camera),
and to support volume rendering.  Using predefined
(or ``special'') strings for {\tt pushVRMLname} makes downstream
ActionScript 3 or JavaScript code {\em universal}; e.g.\ one Adobe 3D
JavaScript code for rotating billboards correctly will work for all
\splot VRML output containing billboards.


X3D was chosen as the most appropriate interchange format for the following
reasons:
\begin{itemize}

\item For the set of features requiring support, X3D is
equivalent to VRML 2.0, easily convertible from VRML, and \splot
was already capable of outputting VRML.

\item XML-encoded X3D is very simple to read and parse in Flash;
ActionScript 3 has XML parsing as a native part of the language, and
the XML Document Object Model (DOM) structure is mirrored by the 
ActionScript 3 objects created.
This is particularly useful for parsing the hierarchical geometry
model tree, and maps isomorphically to the Papervision3D geometry tree.

\item Although text-based X3D documents are larger than
the equivalent geometry expressed in a binary format, Flash can easily
read zipped files.  A zipped archive can also contain the necessary
textures for a scene, so an entire 3-d scene can be exchanged in a
single file (this is much like Google Earth's KMZ
files, which are zipped archives of Collada files with textures).


\end{itemize}

Our final Web-deployment platform is therefore a Flash Papervision3D
application, capable of reading XML-encoded X3D converted output from
\splot programs, and providing a very similar user interface to \splotns.  
This enables two types of applications: a Flash
X3D static geometry-viewer (like the native \splot program \textsc{s2view}),
and custom \splot Flash applications.  The viewer has the advantage
that it can display any supported static geometry output by \splotns,
without having to write any ActionScript 3
code (an unfamiliar language to most astronomers).  Nevertheless, the
\splot layer we have built on top of Papervision3D is simple enough
that scenes {\em can}\/ be animated just by editing a short callback,
in the same manner in which geometry is made dynamic in the \splot
library.

\begin{figure}
\includegraphics[width=3in]{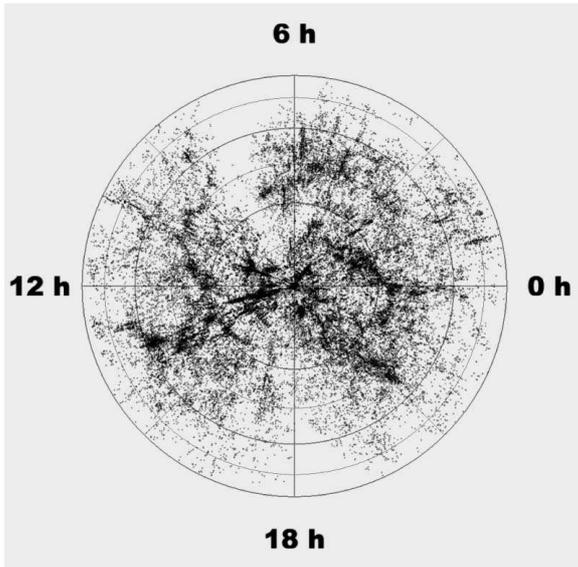} 
\caption{A cone or wedge diagram is the standard representation for
galaxy positions from a redshift survey, usually projected along the line of the
celestial poles. The data is from the 6dF Galaxy
Survey \citep{jones05}.   49~377 galaxy positions, with $0 \leq v \leq$ 15~000 
km s$^{-1}$ and Dec. $\leq 0^\circ$, are shown as black points in this projected
view.  Declination lines are drawn at intervals of $15^\circ$. }
\label{fig:wedge}
\end{figure}

\section{Cosmological Data}
\label{sec:cosmovis}
In order to demonstrate the practicalities of interchanging 
multi-dimensional data, we consider the case of 
cosmological visualization: representation of the three-dimensional 
spatial structure of the Universe, including both observational and 
simulation datasets.   Where such information exists, we extend 
this definition to include time-evolving datasets (e.g. evolution of
structure formation or the hierarchical merging of galaxies), and derived 
data products such as catalogues and merger trees.  A modern, fully-digital 
cosmological visualization allows the user to rotate, zoom, pan and even 
interactively select from datasets.

\subsection{Sources of Data}
Observational cosmological datasets are usually obtained from galaxy
redshift surveys whereby
2-d sky positions (RA, Dec.) are
combined with redshift ($z$) to give a 3-d spatial location.  The
most common representation is the cone or wedge diagram, as shown
in Figure \ref{fig:wedge} using galaxy positions from the 6dF Galaxy Redshift Survey
second data release \citep{jones05}.  An interactive 3-d visualization is 
advantagous in this case, so that the wedge can be rotated and zoomed.
Without interactivity, it is difficult to interpret the details of the 
large-scale structure from a single viewpoint. This conveniently
highlights the limitations of a static, 2-d publishing and presentation 
medium for $n$-dimensional scientific data!


Cosmological simulations generally use $N$-body techniques
to efficiently calculate the gravitational forces between an ensemble
of dark matter, gas and star ``particles''.
Each particle often comes with full information on mass,
spatial position $(x, y, z)$ and velocity $(v_x, v_y, v_z)$ over a range
of time steps (or snapshots).  Derived properties include energy
(kinetic and potential), temperature and density, which may be calculated
on a per particle basis or on a grid.  Dubinski (2008) provides examples 
of standard technqiues for representing N-body data, but with an emphasis 
on pre-rendered sequences rather than real-time interactivity.

\begin{figure*}[t]
\centering
\includegraphics[width=6in]{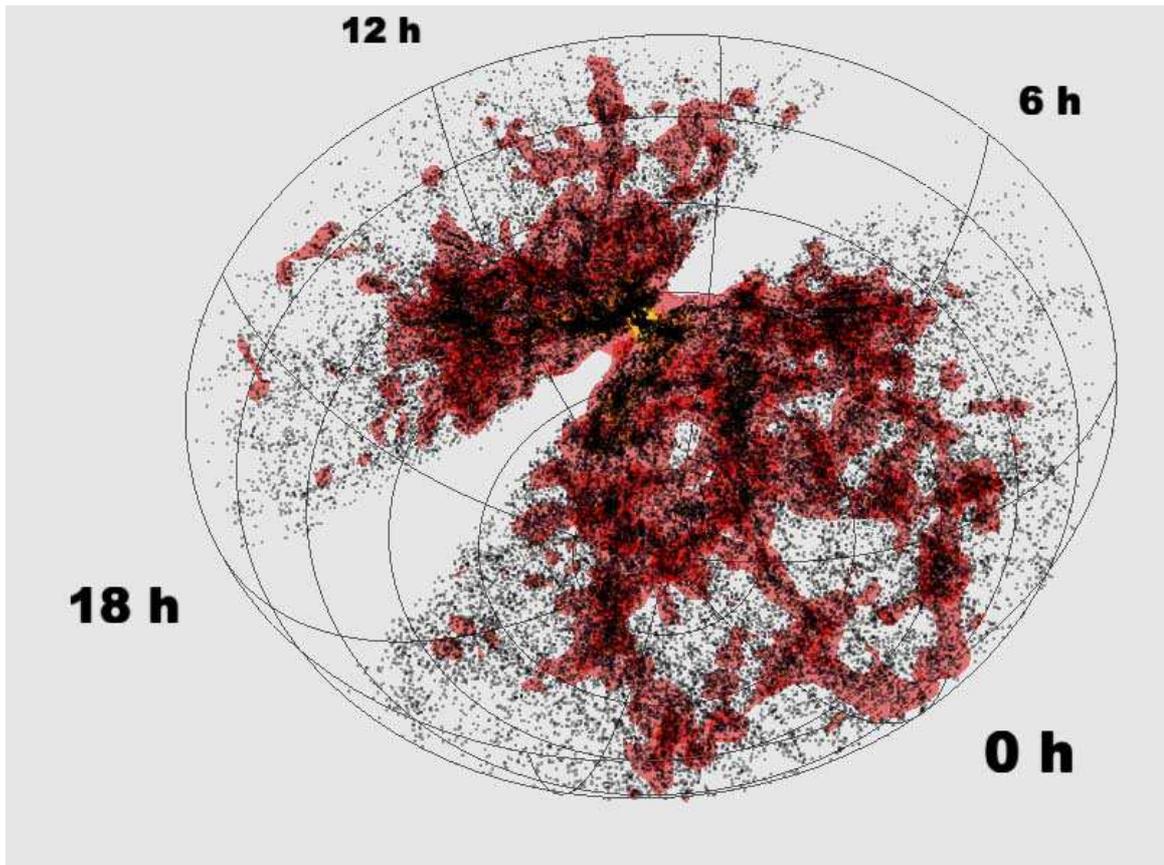} 

\caption{\label{figure:6doblique} Oblique view of the 6dF Galaxy Redshift
Survey (49~377 galaxies with $0 \leq v \leq$ 15~000 km s$^{-1}$ and
Dec. $\leq 0^{\circ}$). Density isosurfaces are overlaid on the galaxy
locations, to help highlight the filamentary large-scale structure.  
Declination lines are shown at intervals of $15^\circ$. }
\end{figure*}

\subsection{Visualization Requirements}
There are four
main data types that a cosmological visualization may need to represent:
\begin{enumerate}
\item {\em 3-d points}: (RA, Dec., $z$) triplets for redshift surveys
and $(x,y,z)$ positions for $N$-body simulations. 
\item {\em 3-d vectors}: ($v_x, v_y, v_z$) velocity data from comological
simulations, which have a spatial location and a direction, or flow lines
representing the motion of particles over time;
\item {\em 2-d and 3-d mesh-based data}: density, temperature, etc. suitable
for display as pixel-based images (2-d) or with volume rendering (3-d). 
Volume rendering (Drebin, Carpenter \& Hanrahan 1988) combines 
gridded data with lighting effects to give
a virtual solid.    Volume rendering requires an opacity to be defined for
each volume element (voxel). In general, this is a combination of absorption and
emission, however, for cosmological datasets we usually only deal with emission: 
$N$-body particles or galaxies are either present or they are not.  
\item {\em non-mesh data}: a practical example is the isodensity surface,
the 3-d analogue of a contour line in 2-d.  These may make 
use of gridded data as a calculation aid, but are not constrained 
to align with the mesh.
\end{enumerate}

The treatment of the first three-dimensions is usally achieved through 
a direct mapping of $(x,y,z)$ positions to screen coordinates for viewing
on a 2-d screen or with a 3-d stereoscopic display. 
Higher data dimensions are visualized through a combination of techniques 
including:
\begin{itemize}
\item {\em colour}:
whether plotting points or mesh data, an appropriate colour map can help
to highlight significant features in a dataset.  Used carelessly, though, colour
is also the best way to ruin the effectiveness of a visualization 
\citep[e.g.][]{tufte90,tufte01};
\item {\em point sizes: } are most suitable when there is a natural
size-ordering (e.g. mass, magnitude), rather than for a morphological 
or descriptive quantity (e.g. galaxy type, survey date).  The latter are 
often better distinguished through the use of symbols or {\em glyphs};
\item {\em splats: } are texture billboards, often with a gaussian
profile, which can be used as an alternative to volume rendering, particularly
when there are a large number of particles \citep{westover89,westover90}.
Splats can demonstrate physical scales by choosing an approprate texture size.
For an example from cosmology, see \citep{hopf03}; and
\item {\em animation:} usually reserved for visualizing variation with time,
although it can also show the variation of a quantity, such as cycling through
a sequence of density isosurfaces.
\end{itemize}

\splot provides support for each of the data types 
and techniques described above, and so is appropriate for generating a range
of cosmological visualizations. To provide 
the requirements of interchange, a move to other output formats requires that 
all of these data types are supported.  Using examples from cosmological 
visualization, we now demonstrate how this has been achieved.

\section{Cosmological Examples}
In this section, we present several specific examples from cosmological 
visualization. We look at how our approach permits interchange between
\splot desktop applications, \textsc{s2slides} presentations and 
\textsc{s2web}.  For 3-d PDF demonstrations of interactive volume rendering 
of a dark matter halo and a wedge diagram based on the CfA2 redshift survey,
we refer the reader to Figure 5 of \citet{barnes08}
and Figure 6 of \citet{fluke08} respectively.

\begin{figure*}
\centering
\includegraphics[width=6in]{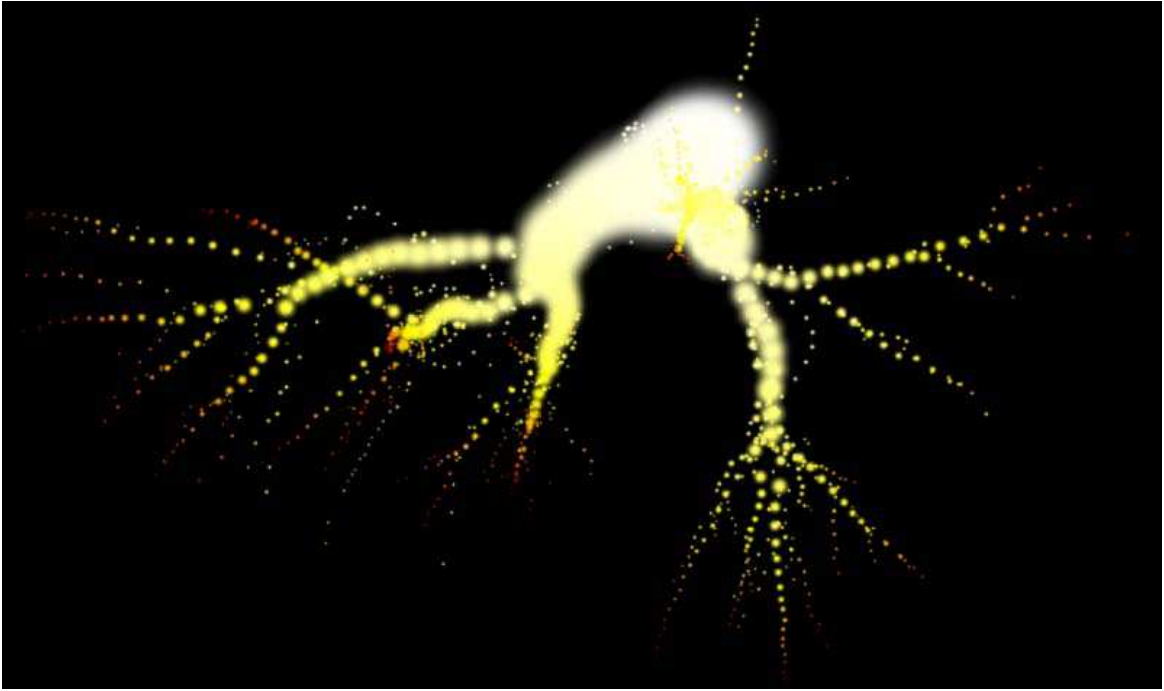} 
\caption{\label{fig:merger} Dark matter halo merger tree from the Millennium Simulation.
Progenitor halo positions are mapped to $(x,y,z)$ coordinates in 3-d space, colour
represents the redshift at which each progenitor is bound to the final halo (``heat'': 
red = high redshift, white = low redshift), texture splat size is scaled by virial radius
of each sub-halo. }
\end{figure*}

\begin{figure*}
\centering
\includegraphics[width=1.9in]{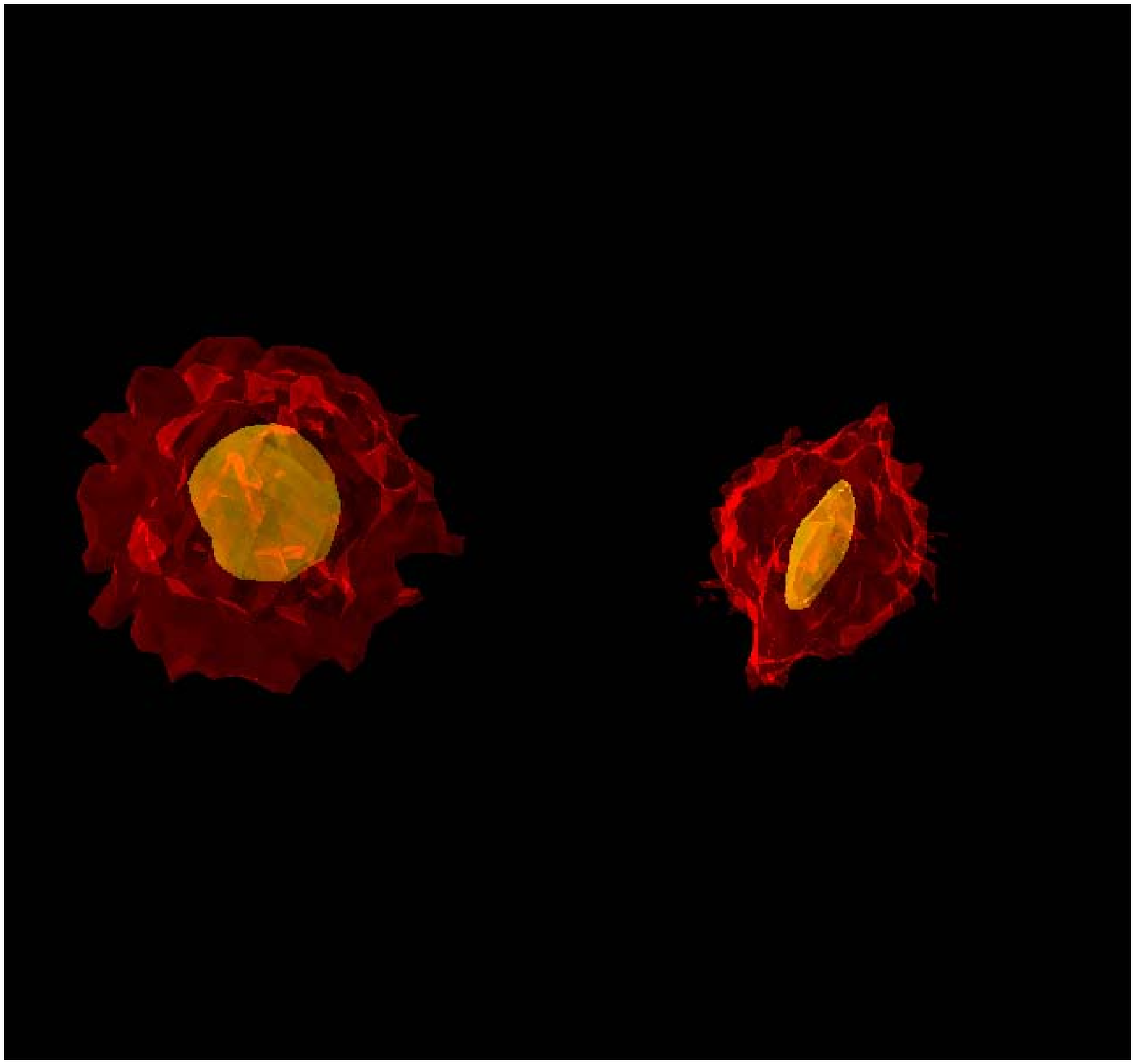} 
\includegraphics[width=1.9in]{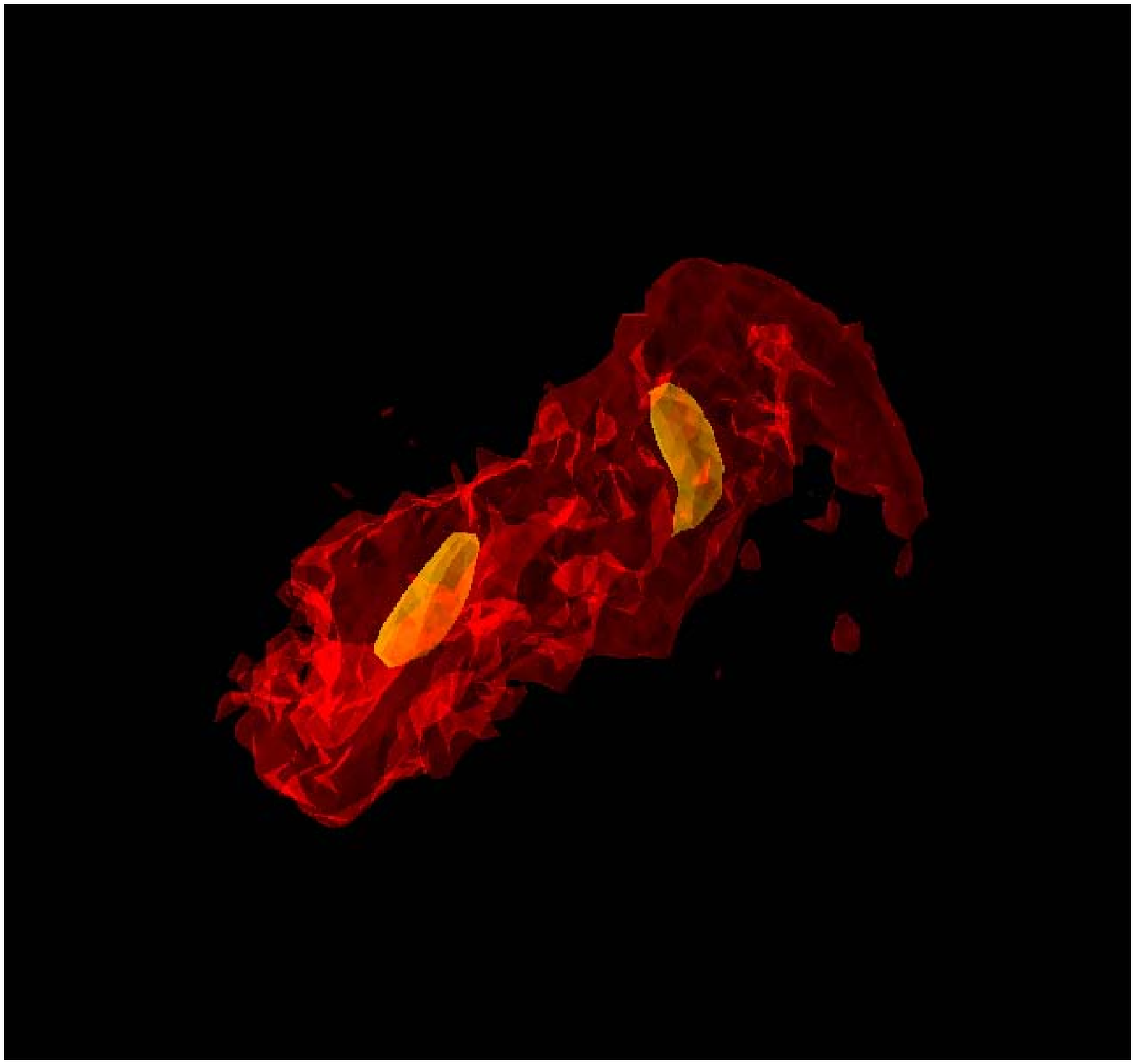} 
\includegraphics[width=1.9in]{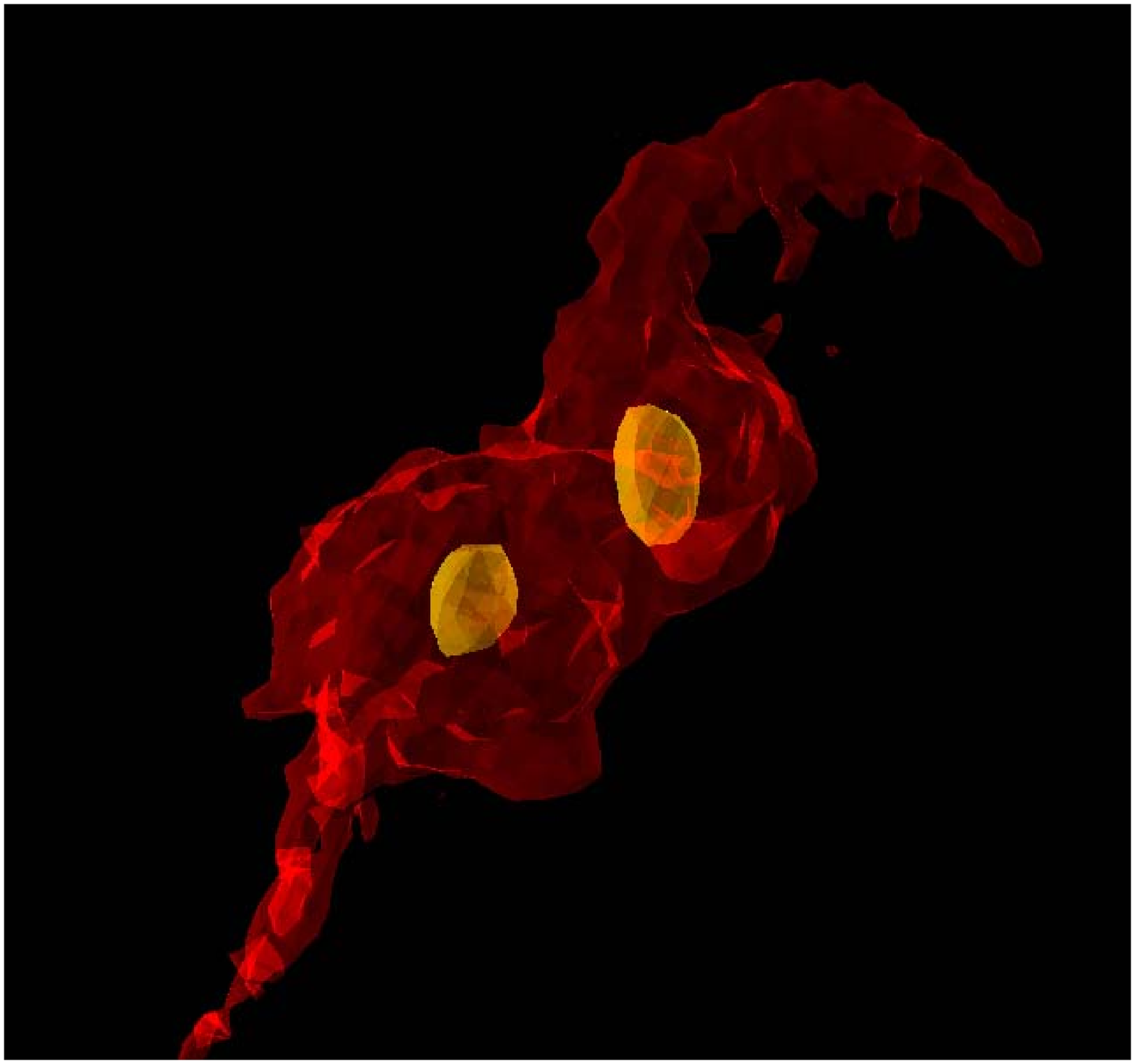} 
\vspace{0.1cm}
\includegraphics[width=1.9in]{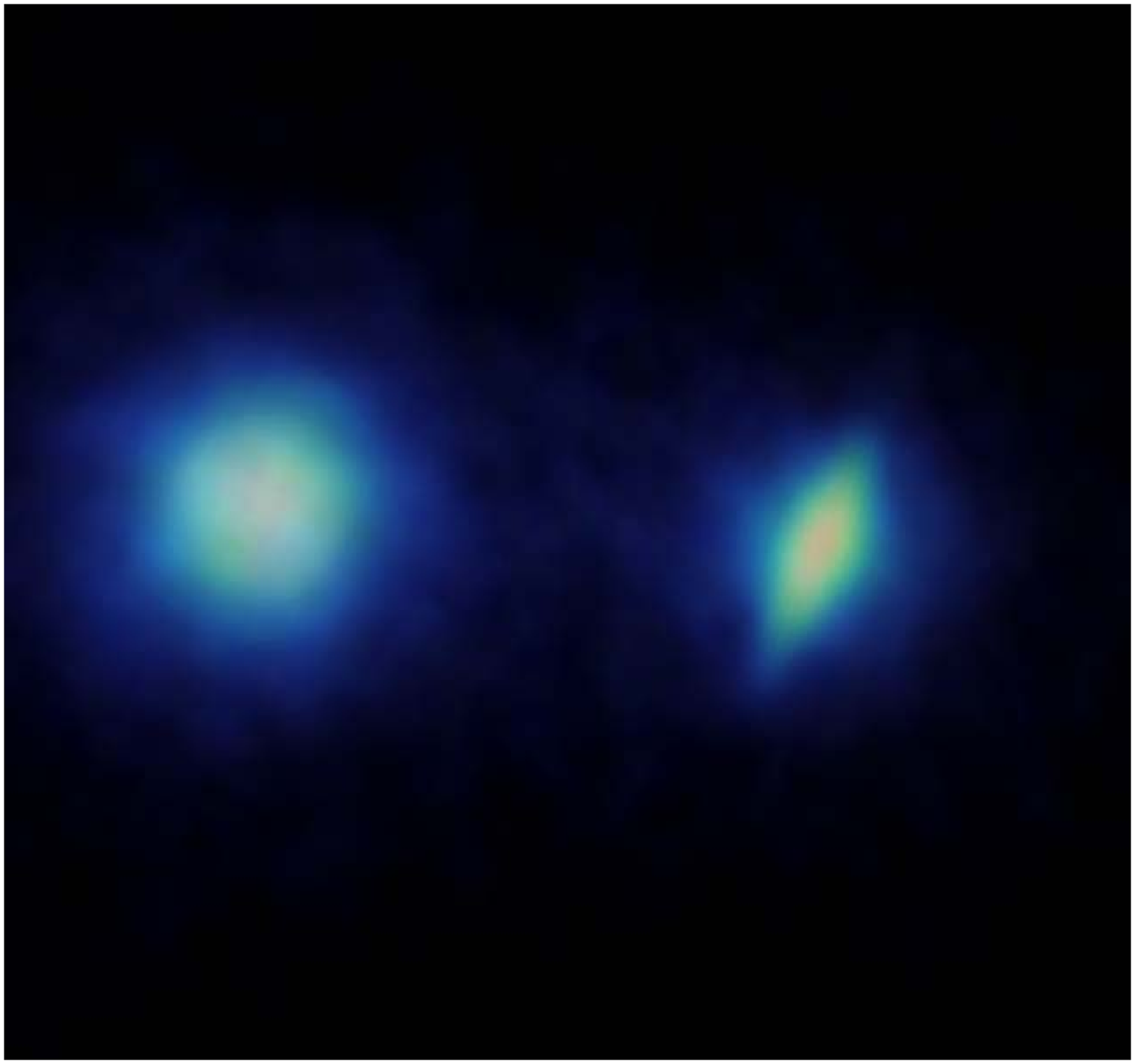} 
\includegraphics[width=1.9in]{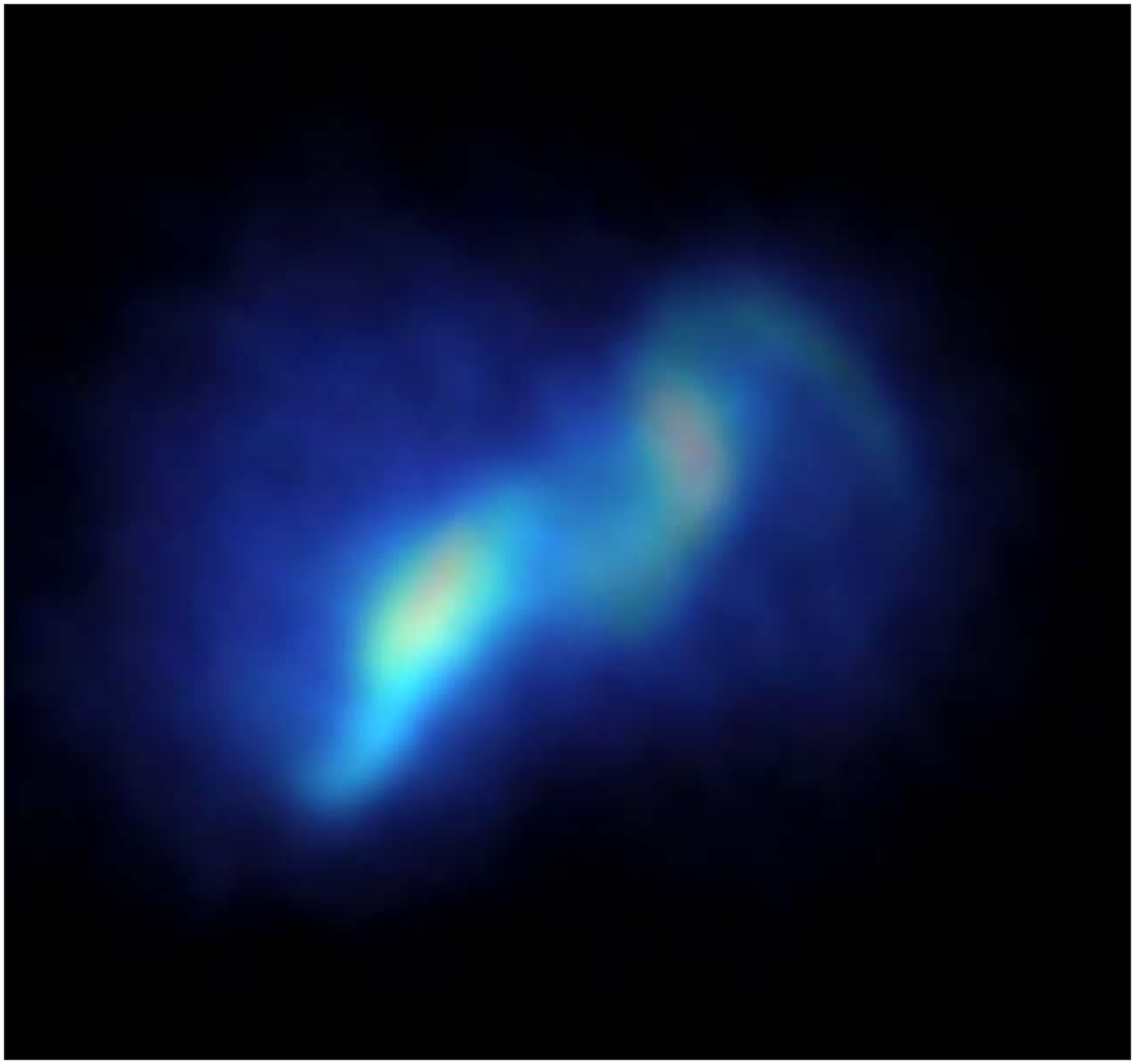} 
\includegraphics[width=1.9in]{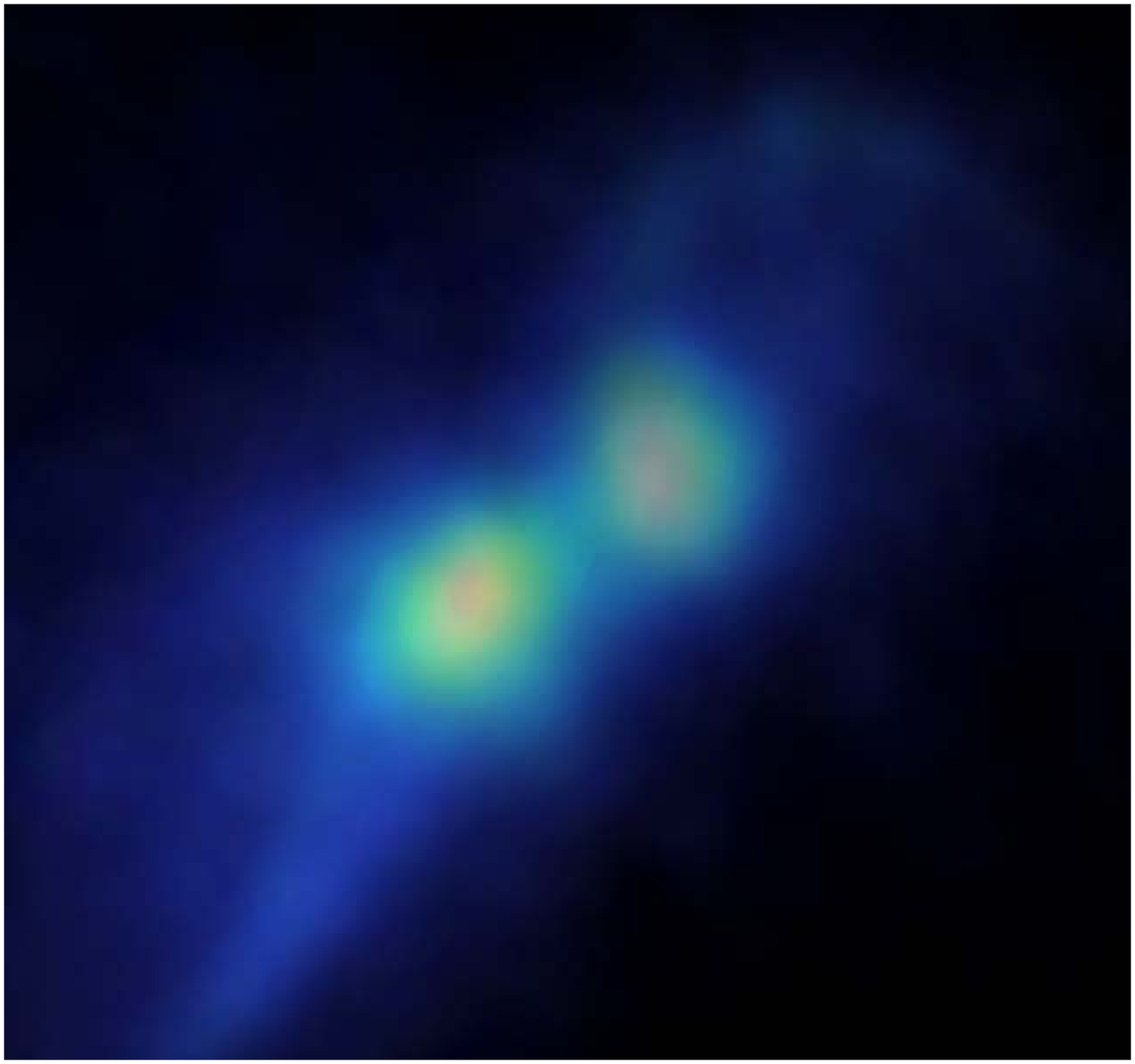} 
\caption{\label{fig:interaction} Three steps in a merger between 
two equal-sized galaxies.  The Gadget-2 simulation comprises 60 000 particles
that are smoothed onto a $100^3$-cell cubic grid using the TSC 
algorithm.  The counts in each cell, $n_{ijk}$, are scaled by 
$\log_{10}(1+n_{ijk})$ and then normalized to lie in the range 0--100.  
(top row)  Density isosurface levels of $20\%$ (red) and $65\%$ (yellow).
(bottom row) Volume rendering with a linear 
``rainbow'' (blue-green-yellow-red from low to high density) colour-map. }
\end{figure*}

\subsection{Wedge Diagram}
As discussed in the previous section, the wedge diagram is the most common
representation of galaxy locations from redshift surveys.  Therefore, it serves
as an important first demonstration of our approach to interchange. 

A static view of the 6dF Galaxy Survey ($0 \leq v \leq$ 15 000 km s$^{-1}$,
Dec. $\leq 0$), projected along the celestial poles, was shown in 
Figure \ref{fig:wedge}.  An oblique view of the same dataset is shown
in Figure \ref{figure:6doblique}, obtained as a screenshot from an 
interactive \splot 
desktop application. The figure shows density isosurfaces overlaid on 
the galaxy locations to help highlight the large-scale filamentary structures.  
The isosurfaces were generated by smoothing galaxy locations
onto an $80^3$-cell cubic mesh using the triangular-shaped cloud algorithm 
(TSC; Hockney \& Eastwood 1988).  The number count, $n_{ijk}$, in each cell 
is rescaled as $\log_{10}(1 + n_{ijk})$, and then linearly normalized to 
lie in the range 0--100.  The final 3-d mesh is converted to an isosurface using
the \splot {\tt ns2cis()} function, and displayed with the {\tt ns2dis()} function.
Both 25\% (red) and 75\% (yellow) contour levels are shown. No attempt has been made 
to correct for the non-uniform volume filling effect of the spherical (RA, Dec., $z$) 
coordinate system when smoothed onto a Cartesian grid in $(x,y,z)$-space.

The VRML output is saved into a file from the \splot program when the
user presses the key combination Shift-W.  The conversion to X3D
format is syntactic only, and we use the freely-available 
``Vrml97toX3dNist'' convertor provided by the National Institute of Standards and
Technology.\footnote{\url{http://ovrt.nist.gov/v2_x3d.html}}
Textures are
written by \splot in TGA format and must be converted to PNG format:
we use the ImageMagick \textsc{convert} 
program.\footnote{\url{http://www.imagemagick.org/}}
The resultant X3D file and PNG texture file(s) are then zipped together into one 
file.  This file can be placed on the Web and an existing \textsc{s2web} 
Flash program pointed to the zip file, or the zip file can be embedded in the Flash 
program itself.  For this paper, we do the latter.

The \textsc{s2web} version of the 6dF Galaxy Survey can be viewed at
\url{http://astronomy.swin.edu.au/s2plot/interchange/6dF}.  Mouse control of camera
orientation, and key presses for functionality such as zooming, toggling autospin, and
changing the spin speed are consistent between the Web and desktop applications.  This
is a critical part of our interchange philosophy -- that there is a common user 
interaction experience.
Of additional interest here is the use of textured billboards for the RA labels, 
which retain their orientation towards the viewer regardless of the camera orientation. 
Line thickness, such as for the grid, is only a global setting
in VRML, and correspondingly X3D.  Therefore \splot programs with
\textsc{s2web} (or PDF) as a target should not use thick lines.  A work-around
is to use thin cylinders instead of lines.

\subsection{Merger Tree}
\label{sct:mergertree}
Semi-analytic models of galaxy formation \citep{white91, cole00, croton06} make 
use of halo merger trees, 
usually generated from dark matter cosmological simulations.  Starting with a catalogue of
halos at $z = 0$, the halo constituents are traced backwards in time within the simulation,
so that the history of mergers can be reconstructed.   Each progenitor halo can then 
be tagged with
a galaxy, and used as an input to a (prescriptive) galaxy formation model.  In this example,
we use data from the Millennium Simulation \citep{springel05b} and the galaxy 
model of \citet{delucia07}.

The \splot application reads a text file generated by an SQL query of the Virgo - Millennium
Database,\footnote{\url{http://www.g-vo.org/Millennium}} returning the list of progenitors
for a given halo.  The application is able to plot any combination of three 
(numerical) parameters in the 3-d space, and then applies a colour map based on the 
fourth parameter.  In Figure \ref{fig:merger}, we show $(x,y,z)$ spatial coordinates for
the progenitor halos, and colour by the redshift at which each progenitor is bound
to the final halo (``heat'' colour map: red = high redshift, white = low redshift).  Rather
than using points to mark each progenitor, we use textured splats, with sizes scaled
by the virial radius of each sub-halo.

Export to VRML and conversion to Web format is the same as for the 6dF case,
and the \textsc{s2web} version is available from 
\url{http://astronomy.swin.edu.au/s2plot/interchange/tree}.  
In this example, some additional steps are required to generate the 
appropriate texture splats.  Within \splotns, a single texture image is
used, and the colour is controlled within the software.
A current limitation in Papervision3D's material attributes for rendering is
that only the first colour applied to the texture will be used.
Accordingly, we post-process the VRML file prior to conversion to X3D,
and where a single texture is used more than once in different
colours, we create multiple copies of the texture so that
Papervision3D and \textsc{s2web} code correctly shades the textures.
Typically, coloured billboards are very small occupying a few tens of
pixels, and so the penalty in total zip file size for this approach is
small.  We intend to build this capability directly into the \textsc{s2web}
code shortly.

\subsection{Galaxy Merger}
\label{sct:animation}
Investigating the dynamics of mergers between galaxies through numerical
simulations has become critical to understanding how galaxies evolve
over ${\cal O}$(Gyr).  Interactive viewing of the galaxy merger shows the 
different ways that such
a system might be seen on the sky. By varying view angle and
merger parameters, such as relative masses of the galaxies or distance
of closest approach, the goal is to reproduce known merging
systems (such as the Antenna galaxies NGC 4038/9, 90 million
light years distant in the constellation Corvus, or the more distant
merging pair NGC 2207 and IC 2163 in Canis Major) and hence
gain insight into their dynamical evolution.

We use here a sample case distributed with 
the Gadget-2 gravity plus smooth-particle hydrodynamics code 
\citep{springel01, springel05a}.  This test case
simulates a merger between two equal-mass spiral galaxies, using
30 000 particles per galaxy.  An \splot application has
been developed to enable interactive visualization of the merger, including
using isosurface (Figure \ref{fig:interaction}; top row) and volume 
rendering (Figure \ref{fig:interaction}; bottom row) techniques to help 
highlight and identify specific features.  
The application can read multiple Gadget-2 datafiles, cycling between 
the series of steps in the merger -- three representive stages in the merger
are shown.  
As with the 6dF wedge diagram, isosurfaces are built by smoothing particle positions
(in this case, onto a $100^3$-cell cubic mesh), and the number counts are log-normalized.
Density isosurface levels of $20\%$ (red) and $65\%$ (yellow) are shown.
A lower resolution mesh ($80^3$) was used for the volume rendering, and a
``rainbow'' colour map (blue-green-yellow-red from low to high density) is applied over
the data range.

Parameters for the visualization were adjusted within the \splot program, and 
the isosurface model was exported to VRML and converted to X3D.  Since PaperVision3D
is a software-only renderer (i.e. not hardware accelerated), 
it is not currently feasible to export the full
resolution geometry for either isosurfaces or volume rendering.   Instead, there is a
trade-off between download speed (due to file size of the geometry), the interaction rate,
and the level of detail displayed. As a compromise, the \textsc{s2web} versions use
a $60^3$-cell mesh for the isosurface calculation and a $40^3$-cell mesh for
the volume rendering.  

An example of using isosurfaces to display a galaxy merger is available at 
\url{http://astronomy.swin.edu.au/s2plot/interchange/iso}.  For this example, we have 
pre-loaded 5 Gadget-2 snapshots. Within the \splot code, these are given unique 
{\tt FRAME*} number identifications for the VRML output -- \textsc{s2web} determines
which of the frames should be displayed, and the user can step through the 
sequence by pressing the spacebar.

A particular application of implicit user-control of the \splot model tree is 
found in axial-slice volume rendering: a 3-d gridded dataset is used to
generate three sets of textures, one set along each lattice axis of
the volume \citep[e.g.][]{lacroute94}.  
Each set is contained within its own branch of the 
model tree (using the predefined {\tt VRSET*} strings with the {\tt pushVRMLname}
function) and then Adobe 3D JavaScript (for PDF) or Adobe ActionScript 3 
(for Flash) code is written which selects the best slice set to draw
for the current camera location and view direction.  
The Flash application can be viewed at:
\url{http://astronomy.swin.edu.au/s2plot/interchange/vr}.

\begin{figure}
\centering
\includegraphics[width=3.0in]{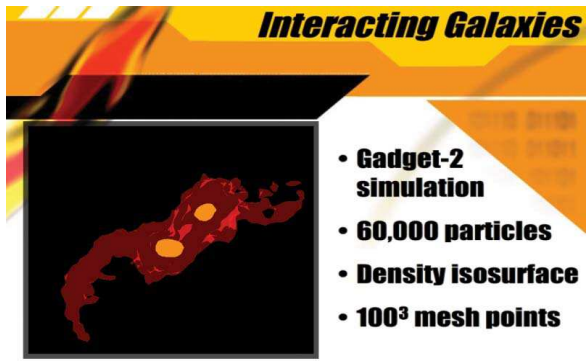} 
\caption{Displaying the interactive view of a Gadget-2 galaxy merger 
via the \textsc{s2slides} application for use in an academic
presentation. The \splot desktop application is overlaid by
\textsc{s2slides} on top of a background image.  The background image
can be created by any image or presentation editor; in this case
Microsoft PowerPoint was used, and the image was exported as a
TGA-format file.}
\label{fig:s2slides}
\end{figure}

Figure \ref{fig:s2slides} shows an example of an \textsc{s2slides} presentation
using the galaxy interaction visualization application.  The same \splot code that is 
used in the desktop application is executed as a child process by 
\textsc{s2slides}.   The position of the application window on the 
background slide is customisable, and a fullscreen mode is available 
via a keystroke.

While Microsoft PowerPoint and Apple Keynote do permit the use of 
embedded interactive datasets on slides through a VRML plug-in, 
they only provide a limited interaction capability, and do not support 
advanced displays.   
One of the main advantages of \textsc{s2slides} is that, as an \splot 
application itself, it can be used with advanced displays.  This reusability means
that a presentation can be given in 2-d via a laptop, but if a stereoscopic
display is available, the presentation can be given in 
stereo simply by changing an environment variable for the display type. 

Some \splot programs may load large data when they run.  In order to 
alert the presenter and audience that something is happening, we utilize 
a spinning logo while applications are starting up.

\section{Future Work}
We have completed most of the work involved in writing the \splot 
model tree structure to VRML 2.0 format, and we have demonstrated most
of the capabilities required for the implementation of a common user
interface across desktop, e-print (PDF) and Web (Flash) targets.  The
latest version of \splot (2.4, October 2008) can export nearly all of
the \splot primitives to VRML, including transparent facets,
billboards and interactive handles.

A binary distribution of the \textsc{s2web} Flash application and a
template HTML file is provided on the \splot website.  We plan to
release the source code for the \textsc{s2web} Flash application in
December 2008.  We will include the code that supports interactive
handles using the {\tt HANDLE*} naming convention in the VRML file;
provide animation capabilities using the {\tt FRAME*} naming
convention; and support axial-slice volume rendering using the {\tt
VRSET*} naming convention.  Users will be able to build the standard
\textsc{s2web} application using the free Adobe Flex 3 SDK, or make simple
modifications to the code for custom animation and interaction.  The
combination of the \splot programming library and the \textsc{s2web} Flash
application yields a free and straightforward technique for
astronomers and other scientists to generate 3-d visualizations on the
desktop and transfer them to the Web.

\begin{figure}
\includegraphics[width=3in]{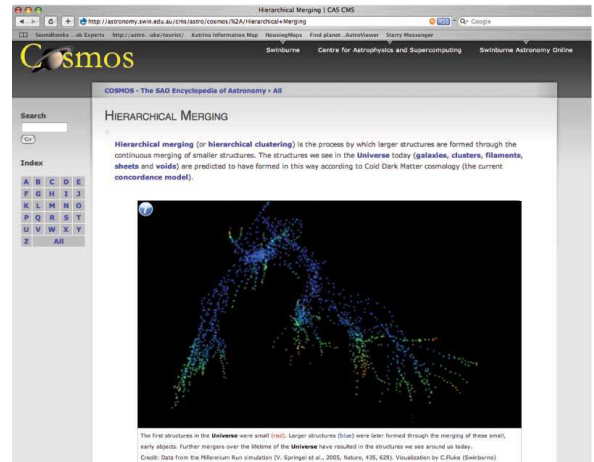} 
\caption{Embedding an interactive merger tree visualization in a COSMOS encyclopedia entry. 
A rainbow colour map is used to indicate progenitor redshifts (red = earliest, blue = latest), 
and there is no scaling of splat sizes. }
\label{fig:hierarchy}
\end{figure}

%

The prospects for future development of the integrated \splot solution
for 3-d analysis, collaboration, publication, education and outreach
are excellent.  Specific areas for improvement and development are
summarized below.

\begin{itemize}

\item Based on Papervision3D and Flash Player 9, \textsc{s2web} is not
hardware accelerated.  Accordingly, its rendering speed, or
equivalently frame rate, depends on both the size of the Flash window
within the Web browser, and the complexity of the 3-d geometry being
displayed.  Volume rendering of data sets larger than around $32^3$
pixels, splatting of more than a few thousand billboards, and the
display of complex transparent isosurfaces, can become slow.  However,
Adobe Flash Player 10 now offers
3-d graphics accelerated by hardware and is presently in beta release.
 It is likely that Papervision3D will be modified to provide hardware
accelerated rendering, thereby improving the usability of \textsc{s2web}
without further work from us.  While we wait for this upgrade though,
we intend to assess the merits of re-writing \textsc{s2web} directly for the
Flash 10 Player, without using the Papervision3D classes.

\item Stereoscopic support: our \splot philosophy has been to support
multiple, advanced display devices from a single library code base.
It should be reasonably straightforward to generate real-time,
stereoscopic 3-d figures within Web browsers, by interleaving the left
and right eye views when a suitable desktop stereo display is
available; and

\item Integration of Web-deployed 3-d assets with our in-house Content
Management System (CMS) to enable straightforward inclusion of 3-d
figures and diagrams in Swinburne Astronomy Online\footnote{SAO:
\url{http://astronomy.swin.edu.au/sao}} course material 
and the COSMOS Encyclopedia of 
Astronomy.\footnote{COSMOS: \url{http://astronomy.swin.edu.au/cosmos/}}  For example, the 
COSMOS entry on hierarchical merging 
(Figure \ref{fig:hierarchy}) contains a version of the 
interactive halo merger tree as described in Section \ref{sct:mergertree}.
With regards to learning and teaching benefits, interactive
presentation and exploration of scientific data promotes student-centred
active learning.  There is a natural opportunity to provide the 
required interactivity via 3-d enabled Web sites, including presentation of
conceptual models, geometrical models, and datasets from both observation
and numerical simulation.  Interactive applications can be
enhanced through features such as dynamics, selection, feedback, etc.,
all of which are feasible with our Flash/ActionScript 3 approach.

\item Our 3-d PDF solution requires commercial software, specifically
Adobe Acrobat 3D or Adobe Acrobat 9 Pro Extended.  We are cogniscent
of the risks of building upon proprietary (and often expensive)
software products, however there are few alternatives at this stage:

\begin{enumerate}

\item We have examined the Universal 3D (U3D) File Format.\footnote{Standard
ECMA-363 Universal 3D File Format:
\url{http://www.ecma-international.org/publications/standards/Ecma-363.htm}}
If we could write suitable U3D files, then these could be embedded in
PDF files as 3-d annotations using the excellent {\tt movie15} \LaTeX\
style created by Alexander Grahn.  Unfortunately there is no good
reference implementation of the U3D standard, and its originators
appear to have lost interest in maintaining or even using the U3D
format.  Furthermore, we have been unable to preserve important
visualization attributes (such as point colour) when experimenting
with U3D files using the Adobe software products.

\item The better format for storing 3-d models in PDF
files is Adobe's PRC format.  We are not aware of any reference
implementations of the PRC format, and the specification has
historically been very hard to locate.  While we cannot write PRC
format directly from \splot at this stage, it may be possible in the
future if Adobe are forthcoming with an open specification of the file
format.  In the meantime, what {\em is}\/ possible is to use an Adobe
product {\em once}\/ to import \splotns-generated VRML, save the
resultant PDF file, and then extract the model data in PRC format
using a tool like {\tt pdftk}.  Once the PRC data is extracted, it can
be easily added to documents generated with {\tt pdflatex} using the
aforementioned {\tt movie15} \LaTeX\ package.

\end{enumerate}

\end{itemize}

\section{Concluding Remarks}
We have described our \splotns-based system for creating 3-d
visualizations, and our extensions which use VRML as an interchange
format to enable the deployment of almost identical 3-d visualizations
across multiple targets: the desktop (\splot application), the
electronic article (3-d PDF), academic presentation (via
\textsc{s2slides}; or alternatively and more portably, using 3-d PDF
documents created with the {\tt beamer} or {\tt prosper}
\LaTeX\ packages), 
and the mainstream Web (Flash \textsc{s2web} application). The two key
standards that our system builds on are the formal 3-d model
interchange format (VRML and X3D) and the informal user interface:
common camera and rendering controls across all output targets.

We have used cosmological visualization as a specific example of the
requirements and benefits of interchange between different target 
modes.  \splot provides the basic set of graphics primitives necessary
to represent wedge diagrams from galaxy redshift surveys and large-scale
structure from cosmological simulations.  Advanced functionality is 
also available, such as volume rendering, splatting and isosurfaces.   We
have demonstrated how cosmological visualization applications built 
on these graphic capablities may be interchanged and accessed interactively
at the desktop, on the Web, and during presentations.   It is evident that
this approach applies equally well across other fields of astronomy research.

While a range of standards-compliant \citep[e.g.][]{wells81,ochsenbein04}
processing, analysis and visualization tools exist, astronomers tend 
to use their own data formats for intermediate steps of processing
(particularly for theory and simulations).  
This avoidance of data standards 
makes interchange of data and intermediate results between collaborators a 
complex process.  It is often only at the completion of a research project 
that astronomers may chose to ``publish'' their data using an 
interchange format, and increasingly this includes Virtual Observatory 
defined standards \citep{norris07}. In most cases, it is easier to 
share still frames or animated sequences from data as intermediate 
products. 

Our approach provides a pathway for the sharing of intermediate and
final data products in an interactive, 3-d format.  The goal here is
that interaction with the Web-based version should be as similar as
possible to the application that was used to explore the data
initially on the researcher's desktop.  Using \textsc{s2web},
interactive models could  be shared online for a ``quick look'', in
the same way that astronomers regularly put images on a personal or
project Web site.

A secondary benefit of interactive visualization in the presentation
and publication modes is the additional level of scrutiny available. 
Rather than selecting a particular ``best'' view for a 2-d still image 
or animation sequence, the viewer is able to explore the dataset for 
themselves, either confirming or rejecting the conclusions as originally 
proposed.   

We believe that 3-d model interchange format(s) need to be augmented
with a common user interface before reasonable uptake
by the community, who have limited time to learn new tools and
techniques, can be expected. The user interface includes interactive
functionality such as the control of viewpoint that has never
been required for classical 2-d graphics, but is axiomatic to 3-d visualization.
By simplifying the deployment of interactive, 3-d visualization
tools that share a common user interface across multiple
media (desktop, e-print, presentation, Web), we aim to properly enable 
and stimulate
the strategic uptake of 3-d visualization techniques in astronomy.

\section*{Acknowledgments} 
This research was supported under Australian Research Council's 
Discovery Projects funding scheme (project number DP0665574).
We acknowledge the Papervision3D development community for
their cohesive work towards a stable and capable 3-d platform for
Flash applications. We thank the anonymous referee for insightful
comments on this paper.  We also thank Alexander Grahn for updating
and sharing his {\tt movie15} \LaTeX\ style file with us, and for
describing how to extract PRC-format model data from PDF files for
re-use with {\tt pdflatex} and {\tt movie15}.


\end{document}